\documentclass[aps,showpacs,twocolumn]{revtex4}
\usepackage{epsfig}
\usepackage[colorlinks=true, letterpaper=true, pdfstartview=FitV, linkcolor=blue, citecolor=blue, urlcolor=blue]{hyperref}
\usepackage{float}

\begin{document}

\title{Investigation of $qqqs\bar q$ pentaquarks in a chiral quark model}

\author{Liting Qin}
\email[E-mail: ]{181002015@stu.njnu.edu.cn}
\author{Yue Tan}
\email[E-mail: ]{181001003@stu.njnu.edu.cn}
\author{Xiaohuang Hu}
\email[E-mail: ]{181002004@stu.njnu.edu.cn}

\author{Jialun Ping}
\email[E-mail: ]{jlping@njnu.edu.cn (corresponding author)}
\affiliation{Department of Physics and Jiangsu Key Laboratory for Numerical
Simulation of Large Scale Complex Systems, Nanjing Normal University, Nanjing 210023, People's Republic of China}

\begin{abstract}
We investigate the pentaquark system  $qqqs \bar q$ in a framework of chiral quark model. Two structures, $(qqq)(s\bar{q})$ and $(qqs)(q\bar{q})$, with all
possible color, spin, flavor configurations are considered. The calculations show that there are several possible resonance states, $Sigma \pi$ and $N \bar{K}$ state with $IJ^P=0\frac{1}{2}^-$,
$\Sigma^* \pi$ with $IJ^P=0\frac{3}{2}^-$,
$\Sigma^* \rho$ with $IJ^P=0\frac{5}{2}^-$,
$\Delta \bar{K}$ with $IJ^P=1\frac{3}{2}^-$ and
$\Delta \bar{K}^*$ with $IJ^P=1\frac{5}{2}^-$.
Where the $N \bar{K}$ state with $IJ^P=0\frac{1}{2}^-$ can be used to explain the $\Lambda(1405)$, and together with another state $\Sigma\pi$ is related to the two-pole structure of the scattering amplitude proposed before. The decay properties of $\Lambda(1520)$ prevent the assignment of $\Sigma^* \pi$ with $IJ^P=0\frac{3}{2}^-$ to $\Lambda(1520)$, although the energy $\sim$ 1518 MeV of $\Sigma^* \pi$ is close to experimental value of $\Lambda(1520)$. Other resonance states generally have a large
width.
\end{abstract}

\maketitle

\section{Introduction}

After decades of development, the quark model has been very successful describing the properties of hadrons. The traditional quark model
believes that there are two types of hadrons in nature, baryons ($qqq$) and mesons ($q \bar q$) respectively. But in addition to their
existence, quantum chromodynamics(QCD) also allows other forms of hadron states such as glueballs (without quarks and antiquarks),
hybrids (gluons mixed with quarks and/or antiquarks), molecular states and compact multiquark states. At present, the low-lying
hadron states can be described well by the traditional quark model. But for the excited states, the traditional quark model encountered
serious problems. For instance, the first excited state of nucleon is expected to be the state with negative parity ($L=1$),
that is $N^*(1535)$, experimentally one has $N^*(1440)$ instead~\cite{PDG}. For the orbital excited state with $L=1$, the mass of $N^*(1535)$
without strangeness should be significantly lower than that of $\Lambda^*(1405)$ with strangeness $-1$ in theory. But the experimental results
are the opposite.

To solve these problems, pentaquark states are proposed. Zou held that the $N^*(1535)$ might be the lowest $L=1$ orbital excited
$|uud\rangle$ state with a large admixture of $|[ud][us]\bar{s} \rangle$ pentaquark component and the $N^*(1440)$ is probably the lowest
radial excited $|uud\rangle$ state with a large component of $|[ud][ud]\bar{d} \rangle$ pentaquark having two $[ud]$ diquarks in the relative
$P$-wave~\cite{EPJA35-325}. Similarly, the lighter $\Lambda^*(1405)$ has a dominant pentaquark component $|[ud][us]\bar{u}\rangle$~\cite{EPJA35-325}.
In fact, the resonance $\Lambda^*(1405)$ was considered as a quasibound molecule state of the $\bar{K}N$ system before the establishment
of quantum chromodynamics~\cite{PR155-1649,PRD18-4187,NPA594-325}. In these two decades, there are still a lot of work devoted on the nature
of $\Lambda^*(1405)$ state. In the framework of the separable potential model the authors confirmed that in the $\pi \Sigma$ mass spectrum
the coupled-channel chiral model produces two poles which can be related to the $\Lambda^*(1405)$ resonance in the complex energy plane~\cite{6}. Based on the the QCD sum rule method, Kisslinger {\em et al.} claimed that the $\Lambda(1405)$ is accordant with being a strange hybrid baryon~\cite{7}. Using the chiral unitary approach, Sekihara {\em et al.} has found that the $\Lambda^*(1405)$ resonant state has bigger spatial radii and softer form factors than those of the baryons, more importantly, the structure is dominated by the $\bar{K}N$ component to a large extent~\cite{8}. Shevchenko calculated the $K^-d$ scattering length by applying newly obtained coupled-channel $\bar{K}N-\pi \Sigma$ potentials with one- and two-pole versions of the $\Lambda^*(1405)$ resonance, and calculations proves that the two results obtained with it are totally separated from each other, therefore, the author prefer to the $\bar{K}N-\pi \Sigma$ interaction models~\cite{9}. Oller {\em et al.} have got an improved theoretical description to calculate the $\Sigma \pi$ event distributions, according to this, they concluded that $\Lambda^*(1405)$ is composed of two resonance states~\cite{10}. Some theorists discussed the spatial structure of the resonance $\Lambda^*(1405)$ state based on the $\bar KN$ molecular picture with the chiral $\bar KN$ potential~\cite{11,12}. However, this resonance state may be obtained not only by two-body channels, but also by multi-body channels~\cite{13}. such as $\bar KNN$~\cite{14,15,16,17,18,19}, $\bar KKN$~\cite{20,21,22}, $\bar K\bar KN$~\cite{23}.

Except the $\Lambda^*(1405)$ state, the nature of its excited state $\Lambda^*(1520)$ is also in controversy. In the Review of Particle
Physics it is a particle marked with four stars~\cite{PDG}. 
In Ref.~\cite{FBS59-113}, the authors calculated the energy of the $S$- and $P$-wave $\Lambda$ family using five sets of parameters in the chiral
quark model, two states, $\Lambda^*(1405)$ and $\Lambda^*(1520)$, cannot be described as three-quark baryons. In chiral unitary approach, a quasi-bound
state of meson-baryon was taken as $\Lambda^*(1520)$~\cite{PRC73-035209}, and the Weinberg compositeness condition shown that the meson-baryon component of
$\Lambda^*(1520)$ was as high as 87\%~\cite{PRC90-025208}. However, the compositeness of $\Lambda^*(1520)$ states was estimated to be $\sim$ 23\% in
Ref.~\cite{PRD92-034011}.

With the accumulation of the experimental data and the improvement of the quark model, it is expected to do a rigorous calculation of 
hadron states based on the quark model. In this work, we systematically investigate the energy spectrum of five-quark state $qqqs\bar{q}, q=u,d$ 
in the framework of the chiral quark model (ChQM), which describing the hadron as well as hadron-hadron interaction successfully~\cite{JPG31-481,RPP68-965},
and a powerful few-body method, the Gaussian expansion method(GEM)~\cite{GEM}, is employed to do the calculation.
The GEM has proven its power in the benchmark test calculation on four-nucleon bound state~\cite{PRC64-044001}. In the present calculation,
two structures, $(qqq)(s\bar{q})$ and $(qqs)(q\bar{q})$, with all possible color, spin, flavor configurations are considered.

The structure of the present paper is organized as follows. In Sec.II the chiral quark model, pentaquark wave functions and GEM are briefly introduced. 
The calculated results and a discussion are presented in Sec.III. The summary of our investigation is given in the last section.

\section{Model and wave function}

The QCD-inspired quark model is one of the main method for studying hadron properties, hadron-hadron interactions and multiquark
states~\cite{PRD80-114023,PRC100-025203,PRD101-054010}. Here, we apply ChQM to five-quark systems with one $s$ quark. The broken SU(3) flavor
symmetry is used in constructing the hamiltonian for the $u,d,s$ system. In this model, the interaction between quark and quark (antiquark) is through
the color confinement $V^{CON}$, the one-gluon exchange (OGE) $V^{OGE}$, the Goldstone boson exchange $V^{\chi}$ $(\chi=\pi,k,\eta)$, as well as the
scalar nonet (the extension of chiral partner $\sigma$ meson) exchange $V^s$ $(s=\sigma,a_0,\kappa,f_0)$. So the Hamiltonian in the present calculation
takes the form~\cite{JPG31-481,RPP68-965},
\begin{widetext}
\begin{eqnarray}
H & = & \sum_{i=1}^{5}\left( m_i+\frac{p^2_i}{2m_i}\right)-T_{CM}
      + \sum_{j>i=1}^{5}\left[ V^{CON}({{\bf r}_{ij}})+V^{OGE}({{\bf r}_{ij}})
      +V^{\chi}({{\bf r}_{ij}})+V^s({{\bf r}_{ij}})\right] , \\
V^{CON}({{\bf r}_{ij}}) & = & \mbox{\boldmath $\lambda$}_i^c\cdot\mbox{\boldmath $\lambda$}_j^c
   \left[-a_c (1-e^{-\mu_cr_{ij}})+\Delta \right] , \\
V^{OGE}({{\bf r}_{ij}}) & = & \frac{1}{4}\alpha_s\mbox{\boldmath $\lambda$}_i^c\cdot\mbox{\boldmath $\lambda$}_j^c
   \left[ \frac{1}{r_{ij}}-\frac{1}{6m_im_j}\mbox{\boldmath $\sigma$}_i\cdot\mbox{\boldmath $\sigma$}_j
   \frac{e^{-r_{ij}/r_0(\mu)}}{r_{ij}r^2_0(\mu)}\right] , ~~~r_0(\mu)=\hat{r}_0/\mu,~~\alpha_{s} =
   \frac{\alpha_{0}}{\ln(\frac{\mu^2+\mu_{0}^2}{\Lambda_{0}^2})}.  \\
V^{\chi}({{\bf r}_{ij}}) & = & v_{\pi}({{\bf r}_{ij}})\sum_{a=1}^{3}
	(\mbox{\boldmath $\lambda$}_i^a\cdot\mbox{\boldmath $\lambda$}_j^a)+v_{K}({{\bf r}_{ij}})\sum_{a=4}^{7}
	(\mbox{\boldmath $\lambda$}_i^a\cdot\mbox{\boldmath $\lambda$}_j^a)+v_{\eta}({{\bf r}_{ij}})
    [\cos\theta_{P}(\mbox{\boldmath $\lambda$}_i^8\cdot\mbox{\boldmath $\lambda$}_j^8)-\sin\theta_{P}
    (\mbox{\boldmath $\lambda$}_i^0\cdot\mbox{\boldmath $\lambda$}_j^0)] ,  \\
v_{\chi}({{\bf r}_{ij}}) & = & \frac{g^2_{ch}}{4\pi}\frac{m^2_\chi}{12m_im_j}
	\frac{\Lambda^2_\chi}{\Lambda^2_\chi-m^2_\chi}m_\chi
	\left[ Y(m_{\chi}r_{ij})-\frac{\Lambda^3_\chi}{m^3_\chi}Y (\Lambda_{\chi}r_{ij}) \right]
	(\mbox{\boldmath $\sigma$}_i\cdot\mbox{\boldmath $\sigma$}_j),    ~~~~\chi=\pi,K,\eta , \\
V^{s}({{\bf r}_{ij}}) & = & v_{\sigma}({{\bf r}_{ij}})
	(\mbox{\boldmath $\lambda$}_i^0\cdot\mbox{\boldmath $\lambda$}_j^0)+v_{a_0}({{\bf r}_{ij}})\sum_{a=1}^{3}
	(\mbox{\boldmath $\lambda$}_i^a\cdot\mbox{\boldmath $\lambda$}_j^a)+v_{\kappa}({{\bf r}_{ij}})\sum_{a=4}^{7}
	(\mbox{\boldmath $\lambda$}_i^a\cdot\mbox{\boldmath $\lambda$}_j^a)+v_{f_0}({{\bf r}_{ij}})
    (\mbox{\boldmath $\lambda$}_i^8\cdot\mbox{\boldmath $\lambda$}_j^8) ,  \\
v_{s}({{\bf r}_{ij}}) & = & -\frac{g^2_{ch}}{4\pi} \frac{\Lambda^2_s}{\Lambda^2_s-m^2_s}m_s
	\left[ Y(m_{s}r_{ij})-\frac{\Lambda_s}{m_s}Y(\Lambda_{s}r_{ij})\right],   ~~~~s=\sigma,a_0,\kappa,f_0
\end{eqnarray}
\end{widetext}
where $T_{CM}$ is the kinetic energy of the center-of mass motion; {\boldmath $\sigma$} represents the SU(2) Pauli matrices;
{\boldmath $\lambda^c$}, {\boldmath $\lambda$} represent the SU(3) color and flavor Gell-Mann matrices respectively;
$\mu$ is the reduced mass between two interacting quarks; $\alpha_s$ denotes the strong coupling constant of one-gluon exchange
and $Y(x)$ is the standard Yukawa functions.

The model parameters which are fixed by fitting the meson and baryon spectra are listed in Table.~\ref{tab:table1}. Because in quark model,
we cannot obtain the satisfying outcome of both meson spectra and baryon spectra via the same set of parameters, two sets of parameters
are employed in the present calculation to test the model dependence of the results.
\begin{table}[h]
\caption{\label{tab:table1} Quark model parameters}
\begin{tabular}{c|cc cc} \hline \hline
 & & set I & set II \\ \hline
Quark masses       &$m_u$=$m_d$ (MeV) &378.49&~~399.05\\
                   &$m_s$ (MeV)       &504.95&~~500.90\\   \hline
                   &$\Lambda_\pi$ (fm$^{-1}$)  &4.20&~~4.20\\
                   &$\Lambda_\eta=\Lambda_K~$ (fm$^{-1}$)      &5.20&~~5.20\\
                   &$m_\pi$ (fm$^{-1}$)  &0.70&~~0.70\\
Goldstone bosons   &$m_K$ (fm$^{-1}$)  &2.51&~~2.51\\
                   &$m_\eta$ (fm$^{-1}$)  &2.77&~~2.77\\
                   &$g^2_{ch}/(4\pi)$  &0.54&~~0.54\\
                   &$\theta_P(^\circ)$  &-15&~~-15\\  \hline
            &$a_c$ (MeV)  &198.73&~~171.85\\
            &$\mu_c$ (fm$^{-1})$  &0.50&~~0.65\\
Confinement &$\Delta$ (MeV)  &85.18&~~62.68\\
            &$\alpha_{uu}$  &0.59&~~0.85\\
            &$\alpha_{us}$  &0.48&~~0.60\\ \hline
                   &$m_\sigma$ (fm$^{-1}$)  &3.42&~~3.42\\
                   &$\Lambda_\sigma$ (fm$^{-1}$)  &4.20&~~4.20\\
scalar nonet       &$\Lambda_{a_0}=\Lambda_\kappa=\Lambda_{f_0}$ (fm$^{-1}$)  &5.20&~~5.20\\
                   &$m_{a_0}=m_\kappa=m_{f_0}$ (fm$^{-1}$)  &4.97&~~4.97\\  \hline
  OGE   &$\hat{r}_0~$(MeV~fm)  &25.32&~~38.04\\  \hline \hline
\end{tabular}
\end{table}

The five-quark states we want to investigate have one $s$ quark and four light quarks, so only the following states are involved:
$N$, $\Lambda$, $\Sigma$, $\Sigma^*$, $\Delta$, $\pi$, $\bar{K}$, $\rho$, $\bar{K}^*$, $\omega$, $\eta$. The calculated masses for these states
are listed in Table.~\ref{tab:table2}.

\begin{table}[h]
\caption{\label{tab:table2} The masses of ground-state baryons and mesons involved in the calculation (unit: Mev).}
\begin{tabular}{ccccccc}
\hline \hline
              & $N$~ &$\Lambda$~ &$\Sigma$~ &$\Sigma^*$~ &$\Delta$  \\ \hline
ChQM (SET I)~ &825~ &1095~ &1201~ &1268~ &1081~ \\
ChQM (SET II)~ &872~ &1206~ &1320~ &1405~ &1176~ \\
PDG~\cite{PDG}~ &939~ &1116~ &1193~ &1385~ &1232~ \\ \hline
&$\pi$~ &$\bar{K}$~ &$\rho$~ &$\bar{K}^*$~ &$\eta$~ &$\omega$  \\ \hline
ChQM (SET I)~ &123~ &535~ &719~ &844~ &516~ &625~ \\
ChQM (SET II)~ &134~ &663~ &788~ &943~ &484~ &665~ \\
PDG~\cite{PDG}~ &140~ &494~ &775~ &892~ &548~ &783~ \\
\hline \hline
\end{tabular}
\end{table}

The wave function of five-quark system is constructed in the following way. First, the five quarks are separated as two clusters, one is a
three-quark cluster, and another is a quark-antiquark cluster. Then, we construct the wave function for each cluster. At last, the wave function
of five-quark system is obtained by coupling the two clusters wave functions and applying the appropriating antisymmetrization operator to the
coupled wavefunction. The quark has four degrees of freedom: orbital, spin, color, and flavor. in the following we construct the wave function
for each degree of freedom.

\noindent (a) The wave function for the orbital part.

There are four relative motions for a five-quark system, the wave function is constructed as
\begin{equation}
\psi_{LM_L}=\left[ \left[ \left[
  \phi_{n_1l_1}(\mbox{\boldmath $\rho$})\phi_{n_2l_2}(\mbox{\boldmath $\lambda$})\right]_{l}
  \phi_{n_3l_3}(\mbox{\boldmath $r$}) \right]_{l^{\prime}}
  \phi_{n_4l_4}(\mbox{\boldmath $R$}) \right]_{LM_L},
\end{equation}
with Jacobi coordinates
\begin{eqnarray}
{\mbox{\boldmath $\rho$}} & = & {\mbox{\boldmath $x$}}_1-{\mbox{\boldmath $x$}}_2, \nonumber \\
{\mbox{\boldmath $\lambda$}} & = & (\frac{{m_1\mbox{\boldmath $x$}}_1+{m_2\mbox{\boldmath $x$}}_2}{m_1+m_2})-{\mbox{\boldmath $x$}}_3,  \nonumber \\
{\mbox{\boldmath $r$}} & = & {\mbox{\boldmath $x$}}_4-{\mbox{\boldmath $x$}}_5, \nonumber \\
{\mbox{\boldmath $R$}} & = & (\frac{{m_1\mbox{\boldmath $x$}}_1+{m_2\mbox{\boldmath $x$}}_2
  +{m_3\mbox{\boldmath $x$}}_3}{m_1+m_2+m_3})
  -(\frac{{m_4\mbox{\boldmath $x$}}_4+{m_5\mbox{\boldmath $x$}}_5}{m_4+m_5}). \nonumber \\
\end{eqnarray}
where $\phi_{n_1l_1}(\mbox{\boldmath $\rho$})$ represents the relative motion wave function between the first and the second quarks,
$\phi_{n_2l_2}(\mbox{\boldmath $\lambda$})$ indicates the relative motion between the center of mass of the quarks 1 and 2 and the third quarks
in the three-quark cluster. Similarly, $\phi_{n_3l_3}(\mbox{\boldmath $r$})$ denotes the relative motion between the fourth and fifth quarks
in the quark-antiquark cluster, and $\phi_{n_4l_4}(\mbox{\boldmath $R$})$ expresses the relative motion between two clusters.

The orbital wave functions of the system are obtained by solving the Schr\"{o}dinger equation with the help of the Gaussian expansion method.
In this method, the radial part of the orbital wave function is expanded by a set of gaussians~\cite{GEM},
\begin{eqnarray}
\psi_{lm}(\mathbf{r})=\sum^{n_{max}}_{n=1}c_{nl}\phi^{G}_{nlm}(\mathbf{r})
\end{eqnarray}
\begin{eqnarray}
\phi^{G}_{nlm}(\mathbf{r})=\emph{N}_{nl}r^{l}e^{-\nu_{n}r^{2}}\emph{Y}_{lm}(\hat{\mathbf{r}})
\end{eqnarray}
\begin{eqnarray}
\emph{N}_{nl}=\left(\frac{2^{l+2}(2\nu_{n})^{l+3/2}}{\sqrt\pi(2l+1)!!}\right)^{\frac{1}{2}},
\end{eqnarray}
where $N_{nl}$ is the normalization constant, and $c_{nl}$ is the variational parameter, which is determined by the dynamics of the system.
The Gaussian size parameters are chosen according to the following geometric progression:
\begin{eqnarray}
\nu_{n}=\frac{1}{r^{2}_{n}}, r_{n}=r_{min}a^{n-1}, a=\left(\frac{r_{max}}{r_{min}}\right)^{\frac{1}{n_{max}-1}},
\end{eqnarray}
where the $n_{max}$ is the number of gaussian functions, which is determined by requiring stability of the results.

\noindent (b) The wave function for the flavor part.

There are two possible separations for a five-quark system containing one $s$ quark, one is $(qqq)(s \bar q)$, and another is $(qqs)(q \bar q)$,
$q=u,d$. The flavor wave functions for the three-quark and quark-antiquark clusters are
\begin{eqnarray}
&& |B_{\frac12,\frac12}^{f1}\rangle =\frac{1}{\sqrt{6}}(2uud-udu-duu), \nonumber \\
&& |B_{\frac12,\frac12}^{f2}\rangle =\frac{1}{\sqrt{2}}(udu-duu), \nonumber \\
&& |B_{\frac12,-\frac12}^{f1}\rangle =\frac{1}{\sqrt{6}}(udd+dud-2ddu), \nonumber \\
&& |B_{\frac12,-\frac12}^{f2}\rangle =\frac{1}{\sqrt{2}}(udd-dud),  \nonumber \\
&& |B_{\frac32,\frac32}^f\rangle = uuu, \nonumber \\
&& |B_{\frac32,\frac12}^f\rangle = \frac{1}{\sqrt{3}}(uud+udu+duu), \nonumber \\
&& |B_{\frac32,-\frac12}^f\rangle = \frac{1}{\sqrt{3}}(udd+dud+ddu), \nonumber \\
&& |B_{\frac32,-\frac32}^f\rangle = ddd, \nonumber \\
&& |B_{0,0}^f\rangle = \frac{1}{\sqrt{2}}(uds-dus), \nonumber \\
&& |B_{1,0}^f\rangle = \frac{1}{\sqrt{2}}(uds+dus), \nonumber \\
&& |B_{1,1}^f\rangle = uus, \nonumber \\
&& |B_{1,-1}^f\rangle = dds, \nonumber \\
&& |M_{\frac12,\frac12}^f\rangle = s \bar d , \nonumber \\
&& |M_{\frac12,-\frac12}^f\rangle = -s \bar u , \nonumber \\
&& |M_{1,0}^f\rangle = \frac{1}{\sqrt{2}}(-u \bar u+d \bar d) , \nonumber \\
&& |M_{1,-1}^f\rangle = -d \bar u , \nonumber \\
&& |M_{1,1}^f\rangle = u \bar d , \nonumber \\
&& |M_{0,0}^f\rangle = \frac{1}{\sqrt{2}}(-u \bar u-d \bar d) .
\end{eqnarray}
The flavor wavefunctions for 5-quark system with isospin $I=0$ are obtained by
the following couplings,
\begin{footnotesize}
\begin{eqnarray}
|\chi^{f1}_{0,0} \rangle & = & \sqrt{\frac{1}{2}}|B_{\frac12,\frac12}^{f1}\rangle |M_{\frac12,-\frac12}^f\rangle-\sqrt{\frac{1}{2}}|B_{\frac12,-\frac12}^{f1}\rangle |M_{\frac12,\frac12}^f\rangle,
\nonumber \\
|\chi^{f2}_{0,0} \rangle & = & \sqrt{\frac{1}{2}}|B_{\frac12,\frac12}^{f2}\rangle |M_{\frac12,-\frac12}^f\rangle-\sqrt{\frac{1}{2}}|B_{\frac12,-\frac12}^{f2}\rangle |M_{\frac12,\frac12}^f\rangle,
\nonumber \\
|\chi^{f3}_{0,0} \rangle & = & \sqrt{\frac{1}{3}}|B_{1,1}^f\rangle |M_{1,-1}^f\rangle-\sqrt{\frac{1}{3}}|B_{1,0}^f\rangle |M_{1,0}^f\rangle+\sqrt{\frac{1}{3}}|B_{1,-1}^f\rangle |M_{1,1}^f\rangle,\nonumber \\
|\chi^{f4}_{0,0} \rangle & = & |B_{0,0}^f\rangle |M_{0,0}^f\rangle. \nonumber \\
\end{eqnarray}
\end{footnotesize}
Similarly, the flavor wavefunctions with isospin $I=1$ are
\begin{footnotesize}
\begin{eqnarray}
|\chi^{f4}_{1,1} \rangle & = & |B_{\frac12,\frac12}^{f1}\rangle |M_{\frac12,\frac12}^f\rangle,
\nonumber \\
|\chi^{f5}_{1,1} \rangle & = & |B_{\frac12,\frac12}^{f2}\rangle |M_{\frac12,\frac12}^f\rangle,
\nonumber \\
|\chi^{f6}_{1,1} \rangle & = & \sqrt{\frac{3}{4}}|B_{\frac32,\frac32}^f\rangle |M_{\frac12,-\frac12}^f\rangle-\sqrt{\frac{1}{4}}|B_{\frac32,\frac12}^f\rangle |M_{\frac12,\frac12}^f\rangle,
\nonumber \\
|\chi^{f7}_{1,1} \rangle & = & |B_{0,0}^{f}\rangle |M_{1,1}^f\rangle,
\nonumber \\
|\chi^{f8}_{1,1} \rangle & = & \sqrt{\frac{1}{2}}|B_{1,1}^f\rangle |M_{1,0}^f\rangle-\sqrt{\frac{1}{2}}|B_{1,0}^f\rangle |M_{1,1}^f\rangle,
\nonumber \\
|\chi^{f9}_{1,1} \rangle & = & |B_{1,1}^f\rangle |M_{0,0}^f\rangle,
\end{eqnarray}
\end{footnotesize}
and the flavor wavefunctions with isospin $I=2$ are
\begin{footnotesize}
\begin{eqnarray}
|\chi^{f9}_{2,2} \rangle & = & |B_{\frac32,\frac32}^{f}\rangle |M_{\frac12,\frac12}^f\rangle,
\nonumber \\
|\chi^{f10}_{2,2} \rangle & = & |B_{1,1}^{f}\rangle |M_{1,1}^f\rangle,
\end{eqnarray}
\end{footnotesize}

\noindent (c) The wave function for the spin part.

In a similar way as the flavor part, the spin wave functions of the three-quark and quark-antiquark clusters are written as,
\begin{eqnarray}
&& |B_{\frac12,\frac12}^{\sigma1}\rangle =\frac{1}{\sqrt{6}}
   (2\alpha\alpha\beta-\alpha\beta\alpha-\beta\alpha\alpha),~~ \nonumber \\
&& |B_{\frac12,\frac12}^{\sigma2}\rangle =\frac{1}{\sqrt{2}}
   (\alpha\beta\alpha-\beta\alpha\alpha), \nonumber \\
&& |B_{\frac12,-\frac12}^{\sigma1}\rangle =\frac{1}{\sqrt{6}}
   (\alpha\beta\beta+\beta\alpha\beta-2\beta\alpha\alpha),~~ \nonumber \\
&& |B_{\frac12,-\frac12}^{\sigma2}\rangle =\frac{1}{\sqrt{2}}
   (\alpha\beta\beta-\beta\alpha\beta), \nonumber \\
&& |B_{\frac32,\frac32}^{\sigma}\rangle =\alpha\alpha\alpha, \nonumber \\
&& |B_{\frac32,\frac12}^{\sigma}\rangle =\frac{1}{\sqrt{3}}
   (\alpha\alpha\beta+\alpha\beta\alpha+\beta\alpha\alpha), \nonumber \\
&& |B_{\frac32,-\frac32}^{\sigma}\rangle =\beta\beta\beta, \nonumber \\
&& |B_{\frac32,-\frac12}^{\sigma}\rangle =\frac{1}{\sqrt{3}}
   (\alpha\beta\beta+\beta\alpha\beta+\beta\beta\alpha), \nonumber \\
&& |M_{1,0}^{\sigma}\rangle =\frac{1}{\sqrt{2}} (\alpha\beta+\beta\alpha). \nonumber \\
&& |M_{1,1}^{\sigma}\rangle =\alpha\alpha. \nonumber \\
&& |M_{1,-1}^{\sigma}\rangle =\beta\beta. \nonumber \\
&& |M_{0,0}^{\sigma}\rangle =\frac{1}{\sqrt{2}} (\alpha\beta-\beta\alpha).
\end{eqnarray}
The spin wavefunctions for 5-quark system with spin $S=\frac12$ are obtained by
the following couplings,
\begin{eqnarray}
|\chi^{\sigma1}_{\frac12,\frac12} \rangle & = & |B_{\frac12,\frac12}^{\sigma1}\rangle |M_{0,0}^{\sigma}\rangle,
\nonumber \\
|\chi^{\sigma2}_{\frac12,\frac12} \rangle & = & |B_{\frac12,\frac12}^{\sigma2}\rangle |M_{0,0}^{\sigma}\rangle,
\nonumber \\
|\chi^{\sigma3}_{\frac12,\frac12} \rangle & = & -\sqrt{\frac{2}{3}}|B_{\frac12,-\frac12}^{\sigma1}\rangle |M_{1,1}^{\sigma}\rangle+\sqrt{\frac{1}{3}}|B_{\frac12,\frac12}^{\sigma1}\rangle |M_{1,0}^{\sigma}\rangle,
\nonumber \\
|\chi^{\sigma4}_{\frac12,\frac12} \rangle & = & -\sqrt{\frac{2}{3}}|B_{\frac12,-\frac12}^{\sigma2}\rangle |M_{1,1}^{\sigma}\rangle+\sqrt{\frac{1}{3}}|B_{\frac12,\frac12}^{\sigma2}\rangle |M_{1,0}^{\sigma}\rangle,
\nonumber \\
|\chi^{\sigma5}_{\frac12,\frac12} \rangle & = & \sqrt{\frac{1}{2}}|B_{\frac32,\frac32}^{\sigma}\rangle |M_{1,-1}^{\sigma}\rangle-\sqrt{\frac{1}{3}}|B_{\frac32,\frac12}^{\sigma}\rangle |M_{1,0}^{\sigma}\rangle
\nonumber \\
&& +\sqrt{\frac{1}{6}}|B_{\frac32,-\frac12}^{\sigma}\rangle |M_{1,1}^{\sigma}\rangle .
\end{eqnarray}
Similarly, the spin wavefunctions with spin $S=\frac32$ are
\begin{eqnarray}
|\chi^{\sigma6}_{\frac32,\frac32} \rangle & = & -|B_{\frac12,\frac12}^{\sigma1}\rangle |M_{1,1}^{\sigma}\rangle,
\nonumber \\
|\chi^{\sigma7}_{\frac32,\frac32} \rangle & = & -|B_{\frac12,\frac12}^{\sigma2}\rangle |M_{1,1}^{\sigma}\rangle,
\nonumber \\
|\chi^{\sigma8}_{\frac32,\frac32} \rangle & = & |B_{\frac32,\frac32}^{\sigma}\rangle |M_{0,0}^{\sigma}\rangle,
\nonumber \\
|\chi^{\sigma9}_{\frac32,\frac32} \rangle & = & \sqrt{\frac{3}{5}}|B_{\frac32,\frac32}^{\sigma}\rangle |M_{1,0}^{\sigma}\rangle-\sqrt{\frac{2}{5}}|B_{\frac32,\frac12}^{\sigma}\rangle |M_{1,1}^{\sigma}\rangle,
\nonumber \\
\end{eqnarray}
and the spin wavefunctions with spin $S=\frac52$ are
\begin{eqnarray}
|\chi^{\sigma10}_{\frac52,\frac52} \rangle & = & |B_{\frac32,\frac32}^{\sigma}\rangle |M_{1,1}^{\sigma}\rangle,
\end{eqnarray}

\noindent (d) The wave function for the color part.

For the color wavefunction, two configurations, color singlet and hidden color are considered. The color wave functions
for two sub-clusters are
\begin{footnotesize}
\begin{eqnarray}
&& |B^{c_1} \rangle = \frac{1}{\sqrt{6}}(rgb-rbg+gbr-grb+brg-bgr), \nonumber \\
&& |B^{c_{2,1}} \rangle = \frac{1}{\sqrt{6}}(2rrg-rgr-grr), ~~~~|B^{c_{2,2}} \rangle = \frac{1}{\sqrt{2}}(rgr-grr), \nonumber \\
&& |B^{c_{3,1}} \rangle = \frac{1}{\sqrt{6}}(rgg+grg-2ggr), ~~~~|B^{c_{3,2}} \rangle = \frac{1}{\sqrt{2}}(rgg-grg), \nonumber \\
&& |B^{c_{4,1}} \rangle = \frac{1}{\sqrt{6}}(2rrb-rbr-brr), ~~~~|B^{c_{4,2}} \rangle = \frac{1}{\sqrt{2}}(rbr-brr), \nonumber \\
&& |B^{c_{5,1}} \rangle = \frac{1}{\sqrt{12}}(2rgb-rbg+2grb-gbr-brg-bgr), \nonumber \\
&& |B^{c_{5,2}} \rangle = \frac{1}{\sqrt{4}}(rbg+gbr-brg-bgr), \nonumber \\
&& |B^{c_{6,1}} \rangle = \frac{1}{\sqrt{12}}(2rgb+rbg-2grb-gbr-brg+bgr), \nonumber \\
&& |B^{c_{6,2}} \rangle = \frac{1}{\sqrt{4}}(rbg-gbr+brg-bgr), \nonumber \\
&& |B^{c_{7,1}} \rangle = \frac{1}{\sqrt{6}}(2ggb-gbg-bgg), ~~~~|B^{c_{7,2}} \rangle = \frac{1}{\sqrt{2}}(gbg-bgg), \nonumber \\
&& |B^{c_{8,1}} \rangle = \frac{1}{\sqrt{6}}(rbb+brb-2bbr), ~~~~|B^{c_{8,2}} \rangle = \frac{1}{\sqrt{2}}(rbb-brb), \nonumber \\
&& |B^{c_{9,1}} \rangle = \frac{1}{\sqrt{6}}(gbb+bgb-2bbg), ~~~~|B^{c_{9,2}} \rangle = \frac{1}{\sqrt{2}}(gbb-bgb), \nonumber \\
&& |M^{c_{1}} \rangle = \frac{1}{\sqrt{3}}(\bar r r+\bar gg+\bar bb), \nonumber \\
&& |M^{c_{2}} \rangle = \bar rb, ~~~~~|M^{c_{3}} \rangle = -\bar gb, ~~~~~|M^{c_{4}} \rangle = -\bar rg\nonumber \\
&& |M^{c_{5}} \rangle = \frac{1}{\sqrt{2}}(\bar r r-\bar gg),~~~~~|M^{c_{6}} \rangle = \frac{1}{\sqrt{6}}(2\bar bb-\bar rr-\bar gg) \nonumber \\
&& |M^{c_{7}} \rangle = -\bar gr, ~~~~~|M^{c_{8}} \rangle = -\bar bg, ~~~~~|M^{c_{9}} \rangle = -\bar br,\nonumber \\
&& |\chi^{c1} \rangle = |B^{c_1} \rangle |M^{c_{1}} \rangle
\nonumber \\
&& ~~~~~~~ =\frac{1}{\sqrt{6}}(rgb-rbg+gbr-grb+brg-bgr)\frac{1}{\sqrt{3}}(\bar r r+\bar gg+\bar bb). \nonumber \\
&& |\chi^{c2} \rangle = \frac{1}{\sqrt{8}}(|B^{c_{2,1}} \rangle |M^{c_{2}} \rangle-|B^{c_{3,1}} \rangle |M^{c_{3}} \rangle-|B^{c_{4,1}} \rangle |M^{c_{4}} \rangle
\nonumber \\
&&~~~~~~~-|B^{c_{7,1}} \rangle |M^{c_{7}} \rangle-|B^{c_{8,1}} \rangle |M^{c_{8}} \rangle+|B^{c_{9,1}} \rangle |M^{c_{9}} \rangle)
\nonumber \\
&& |\chi^{c3} \rangle = \frac{1}{\sqrt{8}}(|B^{c_{2,2}} \rangle |M^{c_{2}} \rangle-|B^{c_{3,2}} \rangle |M^{c_{3}} \rangle-|B^{c_{4,2}} \rangle |M^{c_{4}} \rangle
\nonumber \\
&&~~~~~~~-|B^{c_{7,2}} \rangle |M^{c_{7}} \rangle-|B^{c_{8,2}} \rangle |M^{c_{8}} \rangle+|B^{c_{9,2}} \rangle |M^{c_{9}} \rangle),
\end{eqnarray}
\end{footnotesize}
where $|\chi^{c1} \rangle$ denotes the color singlet configuration, $|\chi^{c2} \rangle$ and $|\chi^{c3} \rangle$ represent the hidden color
configuration.

Finally, the total wave function of the 5-quark system is written as
\begin{eqnarray}
&& \Psi_{JM_J}^{i,j,k}={\cal A} \left[ \left[
 \psi_{L}\chi^{\sigma_i}_{S}\right]_{JM_J}
   \chi^{f}_j \chi^{c}_k \right],\nonumber \\
&&(i=1\sim10,~j=1\sim10,~k=1\sim3),
\end{eqnarray}
where $J$ is the total angular momentum and $M_{J}$ is the 3rd component of the total angular momentum,
and the $\cal{A}$ is the antisymmetry operator of the system, it can be written as
\begin{equation}
 {\cal{A}}=1-(13)-(23)
\end{equation}
for $(qqq)(s \bar q)$ case and
\begin{equation}
 {\cal{A}}=1-(14)-(24)
\end{equation}
for $(qqs)(q\bar q)$ case.
The eigen-energy of the system is obtained by solving the following eigen-equation
\begin{equation}
H\Psi_{JM_J}=E\Psi_{JM_J},
\end{equation}
by using variational principle. The eigen functions $\Psi_{JM_J}$ are the linear combination of the above channel wavefunctions.

\begin{table}
\caption{\label{tab:table3} The possible channels of $(qqq)(s \bar q)$ and $(qqs)(q \bar q)$ systems.}
\begin{ruledtabular}
\begin{tabular}{cc}
$IJ^P$ & channel \\ \hline
$0\frac{1}{2}^-$ & $N \bar{K}$,$N\bar{K}^{*}$,$\Sigma \pi$,$\Sigma \rho$,$\Sigma^* \rho$,$\Lambda \eta$,$\Lambda \omega$ \\
$0\frac{3}{2}^-$ & $N\bar{K}^{*}$,$\Sigma \rho$,$\Sigma^* \pi$,$\Sigma^* \rho$,$\Lambda \omega$ \\
$0\frac{5}{2}^-$ & $\Sigma^* \rho$ \\ \hline
$1\frac{1}{2}^-$ & $N \bar{K}$,$N\bar{K}^{*}$,$\Delta \bar{K}^{*}$,$\Lambda \pi$,$\Lambda \rho$,$\Sigma \pi$,$\Sigma \rho$,$\Sigma^* \rho$,
  $\Sigma \eta$,$\Sigma \omega$,$\Sigma^* \omega$ \\
$1\frac{3}{2}^-$ & $N\bar{K}^{*}$,$\Delta \bar{K}$,$\Delta \bar{K}^{*}$,$\Lambda \rho$,$\Sigma \rho$,$\Sigma^* \pi$,$\Sigma^* \rho$,
  $\Sigma \omega$,$\Sigma^* \eta$,$\Sigma^* \omega$\\
$1\frac{5}{2}^-$ & $\Delta \bar{K}^{*}$,$\Sigma^* \rho$,$\Sigma^* \omega$\\ \hline
$2\frac{1}{2}^-$ & $\Delta \bar{K}^{*}$,$\Sigma \pi$,$\Sigma \rho$,$\Sigma^* \rho$ \\
$2\frac{3}{2}^-$ & $\Delta \bar{K}$,$\Delta \bar{K}^{*}$,$\Sigma \rho$,$\Sigma^* \pi$,$\Sigma^* \rho$\\
$2\frac{5}{2}^-$ & $\Delta \bar{K}^{*}$,$\Sigma^* \rho$\\
\end{tabular}
\end{ruledtabular}
\end{table}

\section{Results and discussions}
\begin{table*}[ht]
\caption{\label{tab:005} The energy of the pentaquark system with $IJ^P=0\frac{1}{2}^-$.
c.c. denotes all color singlet channels coupling.}
\begin{tabular}{ccccccc}
\hline \hline
Index &${c_i \sigma_j f_k}$ & Physical content & $E$ (MeV) & $E^{\rm Theo}_{th}$ (MeV) & $E^{\rm Exp}_{th}$ (MeV) & $E^{\prime}$ (MeV) \\
\hline
1& $i=1;j=1,2;k=1,2$ & $N \bar{K}$ & 1358 & 1362 & 1434 & 1430 \\
2& $i=2,3;j=1,2;k=1,2$ &           & 1933 &      &      &      \\
3& $i=1,2,3;j=1,2;k=1,2$ &         & 1358 &      &      &      \\
4& $i=1;j=3,4;k=1,2$ & $N\bar{K}^{*}$    & 1671 & 1670 & 1831 & 1831  \\
5& $i=2,3;j=3,4;k=1,2$ &           & 1913 &      &      &      \\
6& $i=1,2,3;j=3,4;k=1,2$ &         & 1671 &      &      &      \\
7& $i=1;j=1;k=3$ & $\Sigma \pi$    & 1320 & 1324 & 1329 & 1325  \\
8& $i=2,3;j=1,2;k=3$ &             & 1949 & \\
9& $i=1,2,3;j=1,2;k=3$ &           & 1320 & \\
10& $i=1;j=3;k=3$ & $\Sigma \rho$   & 1923 & 1920 & 1964 & 1964  \\
11& $i=2,3;j=3,4;k=3$ &             & 2405 & \\
12& $i=1,2,3;j=3,4;k=3$ &           & 1923 & \\
13& $i=1;j=5;k=3$ & $\Sigma^* \rho$ & 1990 & 1987 & 2158 & 2158  \\
14& $i=3;j=5;k=3$ &                & 2223 & \\
15& $i=1,3;j=5;k=3$ &              & 1990 & \\
16& $i=1;j=2;k=4$ & $\Lambda \eta$ & 1614 & 1611 & 1664 & 1664  \\
18& $i=2,3;j=1,2;k=4$ &            & 1873 & \\
17& $i=1,2,3;j=1,2;k=4$ &          & 1614 & \\
19& $i=1;j=4;k=4$ &$\Lambda \omega$ & 1724 & 1720 & 1898 & 1898  \\
20& $i=2,3;j=3,4;k=4$ &            &  1978 & \\
21& $i=1,2,3;j=3,4;k=4$ &          &  1724 & \\ \hline
c.c.(SET I) & &    & 1267 & 1324 & 1329 & 1292 \\
            & &    & 1359 & 1362 & 1434 & 1404 \\
c.c.(SET II) & &   & 1396 & 1535 & 1434 & 1282 \\
             & &   & 1505 & 1454 & 1329 & 1389 \\
\hline \hline
\end{tabular}
\end{table*}
In the present work, we try to look for the five-quark systems with quantum numbers $IJ^P(I=0,1,2;J=\frac{1}{2},\frac{3}{2},\frac{5}{2};P=-)$
in the chiral quark model. Two structures, $(qqq)(s \bar q)$ and $(qqs)(q\bar q)$ with color singlet and hidden-color configurations are considered.
We are interested in the low-lying states of the pentaquark systems, so here we set all the orbital angular momenta to zero. Then the parity of
the two configurations of the five-quark system is negative. The possible channels of the two structures are listed in Table.~\ref{tab:table3}.

The calculated results of $IJ^P=0\frac{1}{2}^-$ are given in Table~\ref{tab:005}, where the first column is the index of the channels
involved in the calculation, the second column lists the indices of color, spin and flavor wave functions for every channels, the physical
contents of channels are shown in the third column. The fourth column shows the the calculation results, the fifth and the sixth columns
give the theoretical and experimental thresholds (the sum of the masses of the corresponding baryon and meson), respectively. The last column
shows the corrected energies of the states, which are obtained by
\begin{equation}
E^{\prime}=E+E_{th}^{\rm exp}-E_{th}^{\rm Theo}
\end{equation}
for the single channel calculation. For the results of channel coupling calculation, the corrected energy is defined as
\begin{equation}
E^{\prime}=E+\sum_{i} p_i (E_{th,i}^{\rm exp}-E_{th,i}^{\rm Theo}),
\end{equation}
where $p_i$ is the percentage of the color singlet channel $i$ in the eigen-state.
Due to the chiral quark model cannot give the satisfying outcome of both meson spectra and baryon spectra via the same set of parameters.
By using the corrected energy, we can minimize the systematic error, which appeared in the calculation of the masses of baryons and mesons,
in calculating the energy of pentaquark state. The last four rows show the lowest and next to lowest energies of full color-singlet channels
coupling with two sets of parameters. The hidden-color channels do not affect the low-lying energies because of their high energies compared to the
color singlet channel. The percentages of each color singlet channel in the lowest eigen-state are listed in Table \ref{tab:005a}.
All the results shown in Tables \ref{tab:005} and \ref{tab:005a} are obtained with the first set of parameters.
\begin{table}[h]
\caption{\label{tab:005a} The percentages of color-singlet channels in the lowest and next to lowest eigen-states with $IJ^P=0\frac{1}{2}^-$.}
\begin{tabular}{cccccccccc}
\hline \hline
$E^{\prime}$ (MeV) & $~~N \bar{K}~$ & $~~N\bar{K}^{*}~$ & $~~\Sigma \pi~$ & $~~\Sigma \rho~$ & $~~\Sigma^* \rho~$ & $~~\Lambda \eta~$ & $~~\Lambda \omega~$ \\ \hline
1292 & 29.8\% & 1.8\%  & 66.4\%  & 0.6\%  & 0.1\%  & 0.2\%  & 1.1\%  \\
1404 & 59.1\% & 0.1\%  & 40.3\%  & 0.1\%  & 0.0\%  & 0.1\%  & 0.3\%  \\
\hline \hline
\end{tabular}
\end{table}

In the following we analyze the results in detail.

\begin{table*}[ht]
\caption{\label{tab:006} The energy of the pentaquark system with $IJ^P=0\frac{3}{2}^-$.
c.c. denotes all color singlet channels coupling.}
\begin{tabular}{cccccccccc}
\hline \hline
Index &${c_i \sigma_j f_k}$ & Physical content & $E$ (MeV) & $E^{\rm Theo}_{th}$ (MeV) & $E^{\rm Exp}_{th}$ (MeV) & $E^{\prime}$ (MeV) \\
\hline
 1 & $i=1;j=6,7;k=1,2$     &  $NK^{*}$     & 1664 & 1669 & 1831 & 1826 \\
 2 & $i=2,3;j=6,7;k=1,2$   &               & 1913 &      &      &      \\
 3 & $i=1,2,3;j=6,7;k=1,2$ &               & 1664 &      &      &      \\
 4 & $i=1;j=6;k=3$         & $\Sigma \rho$ & 1919 & 1920 & 1964 & 1963  \\
 5 & $i=2,3;j=6,7;k=1,2$   &               & 2398 &      &      &       \\
 6 & $i=1,2,3;j=6,7;k=1,2$ &               & 1919 &      &      &       \\
 7 & $i=1;j=8;k=3$         & $\Sigma^*\pi$ & 1390 & 1391 & 1523 & 1522  \\
 8 & $i=3;j=8;k=3$         &               & 1987 &      &      &       \\
 9 & $i=1,3;j=8;k=3$       &               & 1390 &      &      &       \\
10 & $i=1;j=9;k=3$         &$\Sigma^*\rho$ & 1989 & 1987 & 2158 & 2158  \\
11 & $i=3;j=9;k=3$         &               & 2123 &      &      &       \\
12 & $i=1,3;j=9;k=3$       &               & 1989 &      &      &       \\
13 & $i=1;j=7;k=4$         &$\Lambda\omega$& 1723 & 1720 & 1898 & 1898  \\
14 & $i=2,3;j=6,7;k=4$     &               & 2297 &      &      &       \\
15 & $i=1,2,3;j=6,7;k=4$   &               & 1723 &      &      &       \\ \hline
c.c.(SET I)  &             &               & 1380 & 1391 & 1523 & 1512  \\
c.c.(SET II) &             &               & 1534 & 1539 & 1523 & 1518  \\
\hline \hline
\end{tabular}
\end{table*}

(a) $IJ^P=0\frac{1}{2}^-$ (Table \ref{tab:005} and Table \ref{tab:005a}): For the color-singlet states, $NK^{*}$, $\Sigma \rho$,
$\Sigma^* \rho$, $\Lambda \eta$, $\Lambda \omega$,
no bound states can be formed in the single-channel calculation, and the energy of the system is almost unchanged by coupling to the
corresponding hidden-color channel. However, we find two bound states in the single-channel calculation, $N \bar{K}$ and $\Sigma \pi$,
with binding energies $-4$ MeV both, the coupling of the corresponding hidden-color channel has no effect on the energy of the system.
So the influence of the hidden color channels on the low-lying states of system can be neglected. This theoretical result is different
from that of Ref.~\cite{33} where the $NK$ state is unbound in the single channel calculation. The reason is that the value of color factor
$\mbox{\boldmath $\lambda$}_i^c\cdot\mbox{\boldmath $\lambda$}_j^c$ for $qq$ is half of that for $q\bar{q}$, and $\pi$ meson exchange potential
is attractive for $u \bar u$ pair and is repulsive for $uu$ pair. The results of all color singlet channels coupling are given in the last
four rows of the Table~\ref{tab:005}. The results show that there is a strong coupling between $N \bar{K}$ and $\Sigma \pi$, the main
component of the lowest state is $\Sigma \pi$, 66.4\%, while the $N \bar{K}$ state takes the percentage 29.8\%. For the next to the lowest
state, the percentages for $\Sigma \pi$ and $N \bar{K}$ are 40.3\% and 59.1\% respectively. The corrected energies of two states are
1292 MeV and 1404 MeV. The state with mass 1404 MeV is naturally taken as candidate of $\Lambda^*(1405)$. Our results can be compared with
that of Ref.~\cite{34}, in which the author put forward two poles of the scattering amplitude between the $N \bar{K}$ and $\Sigma \pi$
thresholds in the complex energy plane to explain the $\Lambda^*(1405)$ resonance state. Two-pole structure of $\Lambda^*(1405)$ was also
claimed in Refs.~\cite{35,36,37,38}.
To check parameter-sensitivity of the results, the second set of parameters is employed to do the calculation. The similar results are obtained,
the lowest state which dominant by $\Sigma \pi$ has has mass 1282 MeV and the second lowest state which dominant by $N \bar{K}$ has the mass 1389 MeV,
10 MeV and 15 MeV away from the value of first set of parameters, respectively.

(b) $IJ^P=0\frac{3}{2}^-$ (Table \ref{tab:006} and Table \ref{tab:006a}): There are three states, $N\bar{K}^{*}$, $\Sigma \rho$ and
$\Sigma^* \pi$ having energy below the corresponding thresholds in the single-channel calculation, the binding energies are
$-5$ MeV, $-1$ MeV and $-1$ MeV, respectively. Similar to the case of $IJ^P=0\frac{1}{2}^-$, coupling to the hidden-color channel does not
change the energies of the states. It is interesting to find herein that after coupling all the color-singlet channels in the $IJ^P=0\frac{3}{2}^-$
system we can get the corrected energy of the lowest state 1512 MeV, which is very close to the experimental mass of $\Lambda^*(1520)$. 
As above, we checked the dependence of the results on the parameters,
the corrected energy of the lowest state is 1518 MeV under the second set of parameters, 6 MeV away from the value of first set of parameters.
However, there is a problem to assign the $\Lambda^*(1520)$ state as the pentaquark state $\Sigma^* \pi$. From Table \ref{tab:006a}, we can see that
the dominant component of the lowest state is $\Sigma^* \pi$, and the partial decay width of $\Sigma^* \pi\rightarrow \Sigma \pi\pi$ is about
3 MeV, which is obtained from decay width of $\Sigma^* \rightarrow \Sigma \pi$, $\sim$4 MeV with phase space correction. But the experimental
value of partial decay width of $\Lambda^*(1520)\rightarrow \Sigma \pi\pi$ is 0.009*15.6=0.14 MeV, which far smaller than 3 MeV.
The fact that the main decay modes of $\Lambda^*(1520)$ are $N\bar{K}$ and $\Sigma \pi$ also support the $3q$ structure of the state
$\Lambda^*(1520)$. Garcia-Recio {\em et al.} studied the compositeness of $\Lambda^*(1520)$, $1-Z=0.227$ also disfavor the baryon-meson explanation
of the state~\cite{PRD92-034011}. Nevertheless, the $\Sigma^* \pi$ as a sizable component of $\Lambda^*(1520)$ is possible when we go beyond the
quenched picture of baryon.

\begin{table}[h]
\caption{\label{tab:006a} The percentages of color-singlet channels in the lowest and next to lowest eigen-states with $IJ^P=0\frac{1}{2}^-$.}
\begin{tabular}{cccccc}
\hline \hline
$E^{\prime}$ (MeV) & $~~N \bar{K}^*~$ & $~~\Sigma \rho~$ & $~~\Sigma^* \pi~$ & $~~\Sigma^* \rho~$ & $~~\Lambda \omega~$ \\ \hline
1512 & 2.3\% & 0.1\%  & 96.5\%  & 0.1\%  & 1.0\%  \\
\hline \hline
\end{tabular}
\end{table}

\begin{table*}
\caption{\label{tab:007} The energy of the pentaquark system with $IJ^P=0\frac{5}{2}^-$.}
\begin{tabular}{ccccccc}
\hline \hline
Index &${c_i \sigma_j f_k}$ & Physical content & $E$ (MeV) & $E^{\rm Theo}_{th}$ (MeV) & $E^{\rm Exp}_{th}$ (MeV) & $E^{\prime}$ (MeV) \\
\hline
1 (SET I) & $i=1;j=10;k=3$   & $\Sigma^{*} \rho$ & 1985 & 1987 & 2160 & 2158 \\
2 (SET I) & $i=3;j=10;k=3$   &                   & 2128 &  &  &  \\
3 (SET I) & $i=1,3;j=10;k=3$ &                   & 1985 &  &  &  \\
1 (SET II)& $i=1;j=10;k=3$   & $\Sigma^{*} \rho$ & 2189 & 2193 & 2160 & 2156 \\
\hline \hline
\end{tabular}
\end{table*}

(c) $IJ^P=0\frac{5}{2}^-$ (Table \ref{tab:007}): In this case there is only one channel $\Sigma^{*} \rho$. The energy of the $\Sigma^{*} \rho$
state obtained is just 2 MeV lower than its threshold, and 4 MeV below the threshold in the calculation with with the second set of parameters.
The hidden-color channel does not change the energy of the system as before. As a result, we can predict it as the pentaquark configuration of
the $\Lambda^*$ with $IJ^P=0\frac{5}{2}^-$. Because of the weak binding, the decay width of the state can be estimated as the sum of $\Sigma^{*}$
decay width and $\rho$ decay width, the state will decay to the $\Lambda \pi \pi \pi$ with the width $\Gamma \sim 185$ MeV.
In PDG~\cite{PDG}, there are two states with masses in the range 2.1$\sim$ 2.2 GeV, $\Lambda(2100) \frac{7}{2}^-$,
$\Lambda(2110) \frac{5}{2}^+$, but the quantum number is a mismatch.

For both systems with $I=1$ and $I=2$, the channel coupling calculation shows that there exists no bound state, so we omit the numerical
results here and just give a brief discussion in the following.

(d) $IJ^P=1\frac{1}{2}^-$: The possible channels are shown in Table \ref{tab:table3}. The single channel calculation cannot find any
bound state, and the channel coupling does not push down any state below the threshold. Two sets of parameters obtain the similar results.
So in this system, no bound states or resonant states may be found.

(e) $IJ^P=1\frac{3}{2}^-$: The single channel calculation reveals that all the states are unbound except the $\Delta \bar{K}$ state
which has binding energy 4 MeV, the corrected energy is 1723 MeV. The lowest energy of the system is 1523 MeV, which is sum of masses
of $\Sigma^*$ and $\pi$. So $\Delta \bar{K}$ may turn out to a resonance state after coupling to $\Sigma^*$ and $\pi$, the dominant decay mode
is $N\bar{K}\pi$ with decay width $\sim$ 120 MeV, which mainly comes from the decay width of $\Delta$. There are a lot of $\Sigma$ states
around 1700 MeV, the states is difficult to be observed experimentally because of its large width.

(f) $IJ^P=1\frac{5}{2}^-$: There are three channels, $\Delta \bar{K}^{*}$, $\Sigma^{*} \rho$ and $\Sigma^{*} \omega$. The single channel
calculation shows that the state $\Delta \bar{K}^{*}$ is bound one with the binding energy of 8 MeV, and other two channels are unbound.
The corrected energy of $\Delta \bar{K}^{*}$ state is 2116 MeV, its decay width is estimated to be $\sim$ 200 MeV. So far there is no
appropriate candidate in PDG~\cite{PDG}.

(g) $IJ^P=2\frac{1}{2}^-$ channel, $IJ^P=2\frac{3}{2}^-$ channel and $IJ^P=2\frac{5}{2}^-$ channel: The results are similar to case (d), there is no
bound state shown up in the single channel calculation and the channel coupling does not help to push down the energy below the threshold.
So there exist no bound state or resonance states with high isospin.

\section{Summary}
In the present work, we investigated the pentaquark state $qqqs \bar q$ in two structures, $(qqq)(s\bar q)$ and $(qqs)(q \bar q)$ based on the
chiral quark model and the Gaussian expansion method. The interesting results are demonstrated in the following: (1) For $IJ^P=0\frac{1}{2}^-$ system, two states are found, one of which is the $\Sigma \pi$ state with the energy of $1282\sim 1292$ MeV and another is the $N \bar{K}$ state with its energy of $1389\sim 1401$ MeV. The results echo the two-pole structure of the scattering amplitude between the $N \bar{K}$ and $\Sigma \pi$ thresholds proposed in explaining the $\Lambda^*(1405)$ resonance state. Particularly, because the energy of the $N \bar{K}$ state is much closer to the $\Lambda(1405)$ state, so we are more inclined to interpret the $\Lambda(1405)$ state as the $N \bar{K}$ state. (2) For $IJ^P=0\frac{3}{2}^-$ system, a resonance state with energy $1512\sim 1518$ MeV is obtained, the main component of which is $\Sigma^{*} \pi$. Although the energy of the state is close to the experimental value of $\Lambda^*(1520)$, the assignment is prevented by the decay properties of $\Lambda^*(1520)$. However, the $\Sigma^{*} \pi$ as a high Fock component of $\Lambda^*(1520)$ is possible. (3) Although in $IJ^P=0\frac{5}{2}^-$ system, there exist only one channel, $\Sigma^* \rho$, it can be a good wide pentaquark resonance with the energy $\sim 2156$ MeV and width $\sim 185$ MeV. (4) For $I=1$ states, $\Delta \bar{K}$ with $J^P=\frac{3}{2}$ and $\Delta \bar{K}^*$ with $J^P=\frac{5}{2}$ are possibly two wide resonance states. Besides, to check the sensitivity of the results to the model parameters, two sets of parameters are employed to perform the calculation, the similar results are obtained.

All calculations in the present work are carried out for the baryon and meson in the ground state. The calculation involved $P$-wave and $D$-wave hadrons will be pursued in the future work.

\section*{Acknowledgments}
The work is supported partly by the National Natural Science Foundation of China under Grant Nos. 11775118, 11535005 and 11675080.

\end{document}